\documentclass[useAMS,usenatbib]{mn2e}
\usepackage{graphicx}
\usepackage{amssymb}

\title[Upper limits on the luminosity of the progenitor of SN2014J]{Upper limits on the luminosity of the progenitor of type Ia supernova SN2014J}
\author[M.T.B. Nielsen, M. Gilfanov, {\'A}, Bogd{\'a}n, Woods, T.E., G. Nelemans]{M.T.B. Nielsen$^{1}$\thanks{E-mail:
mede@mpa-garching.mpg.de}, M. Gilfanov$^{1,2}$, {\'A}. Bogd{\'a}n$^{3}$, T.E. Woods$^{1}$ and G. Nelemans$^{4,5}$ \\
$^{1}$Max-Planck Institut f\"ur Astrophysik, Karl-Schwarzschild-Str. 1, Postfach 1317, D-85741 Garching, Germany\\
$^{2}$Space Research Institute of Russian Academy of Sciences, Profsoyuznaya 84/32,117997 Moscow, Russia\\
$^{3}$Harvard-Smithsonian Center of Astrophysics, 60 Garden Street MS 70, 02138 Cambridge, MA, United States of America \\
$^{4}$Department of Astrophysics, IMAPP, Radboud University Nijmegen, PO Box 9010, NL-6500 GL Nijmegen, the Netherlands\\
$^{5}$Institute for Astronomy, KU Leuven, Celestijnenlaan 200D, 3001 Leuven, Belgium}
\begin{document}

\date{Accepted -. Received \today; in original form -}

\pagerange{\pageref{firstpage}--\pageref{lastpage}} \pubyear{2014}

\maketitle

\label{firstpage}

\begin{abstract}
We analysed  archival data of \textit{Chandra} pre-explosion observations of the position of SN2014J in M82. No X-ray source at this position was detected in the data, and we calculated upper limits on the luminosities of the progenitor. These upper limits allow us to firmly rule out an unobscured supersoft X-ray source progenitor with a photospheric radius comparable to the radius of white dwarf near the Chandrasekhar mass ($\sim$1.38 M$_{\odot}$) and mass accretion rate in the interval where stable nuclear burning can occur. However, due to a relatively large hydrogen column density implied by optical observations of the supernova, we cannot exclude a supersoft source with lower temperatures, $kT \lesssim 80$ eV.  We find that the supernova is located in the center of a large structure of soft diffuse emission, about 200 pc across. The mass, $\sim 3\cdot 10^4$ M$_\odot$ and short cooling time of the gas, $\tau_{\rm cool}\sim 8$ Myrs, suggest that it is a supernova-inflated super-bubble, associated with the region of recent star formation. If SN2014J is indeed located inside the bubble, it likely belongs to the prompt population of type Ia supernovae, with a delay time as short as $\sim 50$ Myrs. Finally, we analysed the one existing post-supernova \textit{Chandra} observation and placed upper limit of $\sim (1-2)\cdot 10^{37}$ erg/s on the X-ray luminosity of the supernova itself. 
\end{abstract}


\section{Introduction} \label{Sect:Introduction}
Type Ia supernovae are highly luminous stellar explosions, visible over cosmological distances. The chemically enriched ejecta from supernova explosions are important to galactic evolution, and may also play a role in the enrichment of the intergalactic medium \citep{Hillebrandt.et.al.2013}. The correlation between peak brightness and exponential fall-off time of the light-curves of type Ia supernovae (\citealt{Philips.1993}) means that they can be used as standardisable cosmological candles to measure the expansion history of the Universe \citep{Riess.et.al.1998,Perlmutter.et.al.1999}. However, the progenitor systems giving rise to these supernovae remain poorly understood, despite decades of studies (see e.g. \citealt{Maoz.et.al.2013}). The consensus is that the exploding objects are carbon-oxygen white dwarfs near the Chandrasekhar mass ($\sim$1.38 M$_{\odot}$) that undergo thermonuclear runaways when the density and pressure in their cores reach values where carbon and oxygen will be processed unstably into radioactive iron-group elements. The resulting explosions completely destroy the white dwarfs, and the subsequent decay of mainly radioactive $^{56}$Ni in the ejecta powers exponential light-curves which, for a period of weeks to months may outshine the integrated luminosity of the host galaxies. However, it remains unclear how a carbon-oxygen white dwarf may reach the Chandraskhar mass. Carbon-oxygen white dwarfs are usually formed at masses significantly lower than that required for thermonuclear runaway, and since there is no known way in which a single white dwarf may grow in mass a binary origin of type Ia supernova progenitors is usually assumed. Two general classes of progenitor scenarios are commonly considered: the single-degenerate and double-degenerate scenarios. In the former, a white dwarf accretes hydrogen-rich material from a non-degenerate companion, and this material is then processed into carbon and oxygen via thermonuclear burning on the surface of the accretor \citep{Whelan.Iben.1973}. In the DD scenario, a binary system goes through two separate white dwarf formation events (or -- in rare cases -- a single event in which both components become white dwarfs simultaneously, see \citealt{Toonen.et.al.2012}), after which the binary loses angular momentum, until the two components merge to form a single white dwarf with a mass at or above that required for a type Ia supernova explosion \citep{Iben.Tutukov.1984,Webbink.1984}. Both scenarios have been studied intensely, but due to the relative scarcity of type Ia supernova events in nearby galaxies direct observational evidence is lacking.

One promising way to distinguish between progenitor scenarios is by X-ray observations. The reason is that the thermonuclear burning of hydrogen-rich material on the surface of the white dwarf in the single-degenerate scenario is expected to be highly luminous in supersoft X-rays. This would mean that single-degenerate progenitors are so-called supersoft X-ray sources (SSSs), characterised by near-Eddington luminosities ($L_{\mathrm{bol}} \sim 4\cdot 10^{37} - 10^{38}$ erg/s) and low effective temperatures ($kT_{\mathrm{eff}}=$ 30-150 eV) \citep{van.den.Heuvel.et.al.1992,Kahabka.van.den.Heuvel.1997}. This SSS state of a single-degenerate progenitor would presumably persist up to the point where the accreting white dwarf explodes as a type Ia supernovae. On the other hand, DD progenitors are not expected to be SSSs \citep{Nielsen.et.al.2014a}. If pre-explosion observations of the positions of known type Ia supernovae can be obtained it may therefore be possible to establish if the progenitors are single-degenerate, accreting systems. As a consequence, a systematic search of the \textit{Chandra} data archive is being undertaken, see \citet{Voss.Nelemans.2008,Roelofs.et.al.2008,Nielsen.et.al.2012,Nielsen.et.al.2013b}, but so far, no unambiguous direct detections of supersoft X-ray emissions from type Ia supernova progenitors have been found. Moreover, the number of observed SSSs and the integrated soft X-ray luminosity of gas-poor galaxies both fall one to two orders of magnitude short of accounting for the expected number of massive, accreting white dwarfs, provided the single-degenerate scenario is the dominant contributor to the type Ia supernova rate \citep{DiStefano.2010,Gilfanov.Bogdan.2010}. This constraint also holds if single-degenerate progenitors radiate at lower temperatures than that typical of most SSSs \citep{Woods.Gilfanov.2013,Johansson.et.al.2014}.

On January 22 2014, the discovery of a new supernova (initially designated PSN\_J09554214+6940260) in M82 was announced and categorised as a young type Ia \citep{Cao.et.al.2014}. Shortly afterwards, it was re-designated SN2014J, and the exact position was given as 09:55:42.121,+69:40:25.88 \citep{STSI.2014}. At a distance of $\sim$3.5 Mpc \citep{Dalcanton.et.al.2009}, this is the closest type Ia supernova since SN1972E more than 40 years ago. Since then, a number of sensitive satellite X-ray telescopes have been in operation, including \textit{EXOSAT}, \textit{Einstein}, \textit{ROSAT}, \textit{XMM-Newton} and \textit{Chandra}. As a result, large amounts of archival X-ray data are now available. The relatively short distance to M82 raised the possibility that SN2014J might yield much-needed, strong observational constraints on the progenitor problem. 

As a continuation of the systematic archival search mentioned above, the authors analysed the publicly available pre-explosion \textit{Chandra} observations of the position of SN2014J to establish if an X-ray source was present. Additionally, we analysed the single available post-explosion observation of SN2014J to establish upper limits on the X-ray luminosity of the supernova itself. In Section \ref{Sect:Data.Reduction} we presents our data analysis, and Section \ref{Sect:Discussion} disusses the implications of our results. Section \ref{Sect:Conclusion} concludes. This article represents an elaboration on the preliminary results reported by \citet{Nielsen.et.al.2014b}, see also \citet{Maksym.et.al.2014}

\section{Data Reduction \& Results} \label{Sect:Data.Reduction}
The \textit{Chandra} data archive contains 22 epochs of publicly available pre-explosion observations covering the position of SN2014J, taken with the CCD Imaging Spectrometer detectors, ACIS-I and ACIS-S. Combined, these epochs amount to 828 ks of observations. In all observations, the supernova position is within an off-axis angle of $\la 3\arcmin$, and for $\approx 43\%$ of the observing time within an off-axis angle of $\la 1\arcmin$. The observations used in this study are summarised in Table \ref{Table:Observations}.

In addition, a post-explosion observation (observation 16580, lasting 46.85 ks) was completed on February 4, 2014, shortly after maximum light of SN2014J (PI R. Margutti). Unlike regular \textit{Chandra} observations which usually become public a year after completion of the observation, this data was made public immediately. The results of their initial analysis have been reported by  \citet{Margutti.et.al.2014}.

\begin{table*}
 \begin{minipage}{140mm}
 \caption{\textit{Chandra} observations used in this study.}
 \centering
  \begin{tabular}{@{}c c c c c c @{}}
  \hline
  \textit{Chandra}	& exposure	& pointing  			& Detector	& observation \\
      observation	& time		& (RA, DEC)			&		& date \\
			& [ks]		&				&		& \\
  \hline
  \hline
    361			& 33.25		& (09:55:51.10,+69:40:45.00)	& ACIS-I	& 1999-09-20 \\
    378			& 4.12		& (09:55:47.00,+69:40:58.00)	& ACIS-I	& 1999-12-30 \\
    379			& 8.94		& (09:55:47.00,+69:40:58.00)	& ACIS-I	& 2000-03-11 \\
    380			& 5.0		& (09:55:47.00,+69:40:58.00)	& ACIS-I	& 2000-05-07 \\
    1302		& 15.52		& (09:55:51.10,+69:40:45.00)	& ACIS-I	& 1999-09-20 \\
    2933		& 18.02		& (09:55:52.60,+69:40:47.10)	& ACIS-S	& 2002-06-18 \\
    5644		& 68.14		& (09:55:50.20,+69:40:47.00)	& ACIS-S	& 2005-08-17 \\
    6097		& 52.77		& (09:55:50.20,+69:40:47.00)	& ACIS-S	& 2005-02-04 \\
    6361		& 17.45		& (09:55:50.20,+69:40:47.00)	& ACIS-S	& 2005-08-18 \\
    8190		& 52.77		& (09:55:50.20,+69:40:47.00)	& ACIS-S	& 2007-06-02 \\
    10025		& 17.4		& (09:55:50.20,+69:40:47.00)	& ACIS-S	& 2009-04-17 \\
    10026		& 16.94		& (09:55:50.20,+69:40:47.00)	& ACIS-S 	& 2009-04-29 \\
    10027		& 18.28		& (09:55:50.20,+69:40:47.00)	& ACIS-S 	& 2008-10-04 \\
    10542		& 118.61	& (09:55:51.30,+69:42:51.60)	& ACIS-S	& 2009-06-24 \\	
    10543		& 118.45	& (09:55:37.60,+69:42:25.10)	& ACIS-S	& 2009-07-01 \\
    10544		& 73.53		& (09:55:54.20,+69:38:57.70)	& ACIS-S	& 2009-07-07 \\
    10545		& 95.04		& (09:56:07.80,+69:39:34.10)	& ACIS-S	& 2010-07-28 \\
    10925		& 44.54		& (09:55:54.20,+69:38:57.70)	& ACIS-S	& 2009-07-07 \\
    11104		& 9.92		& (09:55:46.60,+69:40:38.10)	& ACIS-S	& 2010-06-17 \\
    11800		& 16.82		& (09:56:07.80,+69:39:34.10)	& ACIS-S	& 2010-07-20 \\
    13796		& 19.81		& (09:55:46.60,+69:40:38.10)	& ACIS-S	& 2012-08-09 \\
    15616		& 2.04		& (09:55:50.50,+69:40:40.00)	& ACIS-S	& 2013-02-24 \\
    \hline
\end{tabular} \label{Table:Observations}
\end{minipage}
\end{table*}

For data reduction we used the \textsc{ciao}-4.6 software suite. The analysis in the following expands on that used in \citet{Nielsen.et.al.2012,Nielsen.et.al.2013b}.

As can be seen from Table \ref{Table:Observations}, the observations were spread over a long time interval, the earliest being from shortly after the launch of \textit{Chandra} and the latest little less than a year before the supernova. All observations were reprocessed with \textsc{ciao}'s {\tt chandra\_repro} script. We used the re-calibrated event and support files for the rest of our data analysis.

Initial inspection of the individual observations revealed several observations to have streaks of photons originating from the read-out of a saturated source nearby (designated CXO-M82-J09 in the \textit{Chandra} catalogue). In observation 10542 (the longest exposure of the 22 epochs) one of these streaks crosses the position of the progenitor, hence contaminating it. We removed the streak from observation 10542 using ciao's {\tt acisreadoutcorr} function, and used the corrected observation for the remainder of our data analysis. Other similar streaks are visible in other observations (two such streaks can be seen faintly below the source region on Figure \ref{Fig:0.5-2keV_bkg_diffuse_reg}), but since they cross neither our source nor background regions used in our analysis we did not remove them.

\begin{figure}
 \centering
  \includegraphics[width=1.0\linewidth]{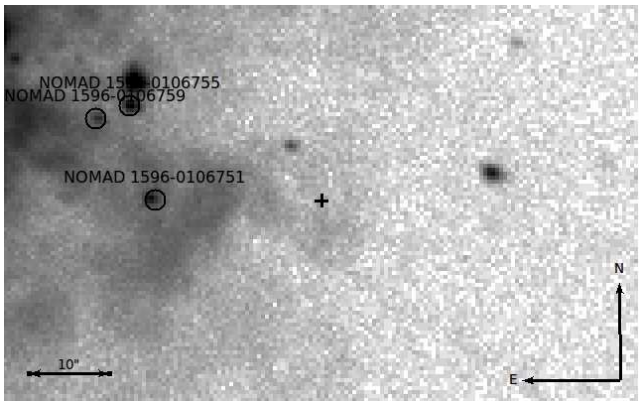}
  \caption{Comparison between optical sources (marked with circles 2.5 arcseconds in diameter) from the Naval Observatory Merged Astrometric Dataset (NOMAD) Catalog (source designations: 1596-0106751, 1596-0106755 \& 1596-0106759) and the stacked image used in our analysis. All three sources are also found in the 2MASS Point Source Catalog, and the source designated 1596-0106759 in NOMAD is additionally found in SDSS6. The cross marks the position of SN2014J.}\label{Fig:astrometry}
\end{figure}

The 22 epochs of pre-explosion observations (with the corrected version of observation 10542) were stacked into one image. To ensure the correct alignment of the Chandra images used in our analysis, we compared prominent point sources in the stacked image with sources in the ds9 catalogues. Figure \ref{Fig:astrometry} shows how the point sources in the stacked image line up with known optical sources, confirming that the astrometry of the image is sufficiently accurate for our purpose.

\subsection{X-ray environment of the progenitor}
\begin{figure}
 \centering
  \includegraphics[width=1.0\linewidth]{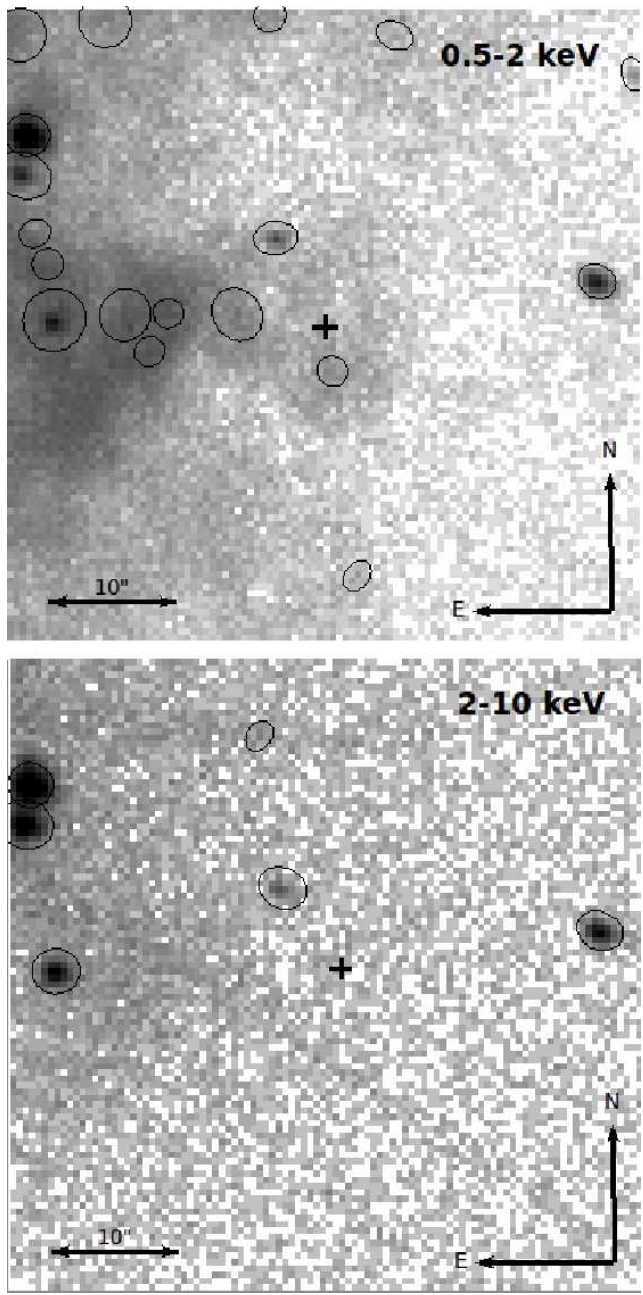} 
  \caption{Part of the stacked image of \textit{Chandra} observations (see Table \ref{Table:Observations}), showing the sources found by {\tt wavdetect} for photon energies in the soft (top) and hard (bottom) bands. The cross marks the position of SN2014J.}\label{Fig:soft_and_hard_energies_wavdetect}
\end{figure}

We filtered the stacked image into a soft (500 eV to 2 keV) and hard (2 to 10 keV) energy band. In the soft band, \textsc{ciao}'s {\tt wavdetect} script finds a point source approximately 7 pixels ($\sim$3.5 arc seconds) South of the reported supernova position, and two other sources within approximately 13 and 15 pixels ($\sim$6.5 and $\sim$7.5 arc seconds) to the East and North-East, as shown in Figure \ref{Fig:soft_and_hard_energies_wavdetect}. These are all too far away to be associated with the supernova progenitor, but could contaminate our measurements if not removed. The source North-East of the supernova progenitor position is also detected in the hard band.

\begin{figure}
 \centering
  \includegraphics[width=1.0\linewidth]{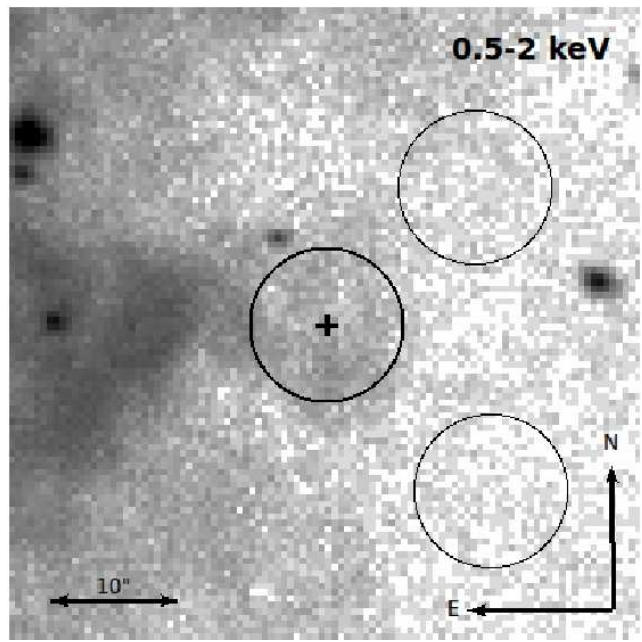} 
  \caption{Part of the stacked image of \textit{Chandra} observations (see Table \ref{Table:Observations}), showing the source region (thick-lined circle, 6 arcsecond radius) and two background regions (thin-lined circle, 6 arcsecond radius) used to extract the spectrum of the region of enhanced diffuse emission in which the supernova is located. The cross marks the position of SN2014J.}\label{Fig:0.5-2keV_bkg_diffuse_reg}
\end{figure}

As can be seen in Figure \ref{Fig:soft_and_hard_energies_wavdetect}, the supernova is located near the center of a roughly circular region of excess soft X-ray  emission, of approximately $\sim 6-8$ arcseconds radius ($\sim$100-140 pc radius, assuming a distance of 3.5 Mpc to M82). To investigate the origin of this emission and its possible connection to the supernova, we extracted a spectrum of a circular region of 6 arcsecond radius centred at the supernova position. We used suitably chosen background regions in the empty areas North and South-East of the soft excess (Figure \ref{Fig:0.5-2keV_bkg_diffuse_reg}). The spectrum was fit with the thermal emission of an optically thin plasma in collisional equilibrium, for which we used \textsc{xspec}'s {\tt mekal} model \citep{Kaastra.Mewe.1993,Liedahl.et.al.1995} with the interstellar absorption modeled by {\tt phabs}. Fixing element abundances at solar values, we obtained the best fit temperature of $kT=0.61\pm 0.03$ keV and the neutral hydrogen column density of $N_{\mathrm{H}}=(8.6\pm 0.4)\cdot 10^{21}$ cm$^{-2}$. Assuming spherical symmetry, from the the model normalization we compute the mean gas density of $n_H\approx 0.19$ cm$^{-3}$ and the total mass of the gas $M_{\rm gas}\approx 3\cdot 10^4$ M$_\odot$. The cooling time of gas equals $\tau_{\rm cool}\approx 8.4$ Myrs. Remembering that M82 has experienced recent star formation, we speculate that the structure could be a super-bubble blown by the stellar winds and core-collapse supernovae in a region of recent star formation. The likely association with massive stars and the short cooling time of the gas suggests that the age of the bubble does not significantly exceed $\sim 40$ Myrs, the lifetime of $\sim 8$ M$_\odot$ stars --  the least massive stars capable of producing core-collapse supernovae \citep{McCray.Kafatos.1987,Mac.Low.McCray.1988}. On the other hand, if SN2014J is indeed located inside the bubble and is a descendant of the same population of young stars, the age of the population should at least exceed $\sim 40$ Myrs, to allow sufficient time for the first white dwarfs to be produced in the population. In order to satisfy both constrains, the age of the stellar population in this region should be around $\sim 40-50$ Myrs and the supernova would then belong to the prompt population of type Ia supernovae, likely associated with massive stars \citep{Mannucci.et.al.2005,Scannapieco.Bildsten.2005}.

In the 2--10 keV hard band, there is no sign of any sources near the progenitor region, besides the one to the North-East that is also present in the soft band. However, in the  6-10 keV band {\tt wavdetect} finds a source located $\sim$1.5 arcsec to the north of the supernova position. The source is associated with a somewhat extended structure located near the supernova position.  Because of this structure, there is a  $\approx 2.4\sigma$ excess above the local background level in the $4.5$ pixel aperture centered at the supernova position. However, the excess virtually disappeared when we excluded the observation \#10542 affected by the readout streak from the analysis. We therefore conclude that the hard source and the excess counts in the supernova position are most likely caused by a somewhat incomplete removal of the readout streak in the observation \#10542. We note that its effect is insignificant in other bands, due to the hard spectrum of the source producing the readout streak.

\subsection{Upper limits on the supernova progenitor luminosity}
In calculating upper limits on the luminosity of a possible SSS at the position of SN2014J, we limited ourselves to photon energies between 300 eV and 1 keV. The reason for adopting 300 eV as a lower limit is the fact that \textit{Chandra}'s response is known to be unreliable for photons below this energy. The upper limit of our energy filtering comes from the expectation that SSSs will not emit any significant amount of energy above 1 keV (e.g. \citealt{Greiner.et.al.1991}).

\begin{figure}
 \centering
  \includegraphics[width=1.0\linewidth]{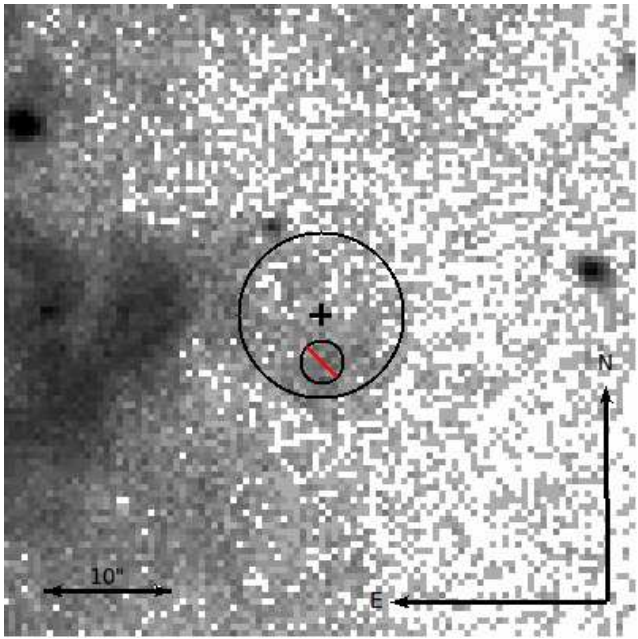}
  \caption{Part of the stacked image of \textit{Chandra} observations (see Table \ref{Table:Observations}), showing the placement of the region from which we extracted the background counts. The excluded, crossed-out circle corresponds to the point source found by {\tt wavdetect} and does not contribute to the background counts. The cross marks the position of the supernova progenitor.}\label{Fig:erange_300:1000}
\end{figure}

We ran {\tt wavdetect} on the 300 eV to 1 keV image and found no sources other than those detected in the 0.5--2 keV band image (Figure \ref{Fig:soft_and_hard_energies_wavdetect}). In order to compute upper limit on the source luminosity we performed standard aperture photometry analysis. We fixed the source region radius at  4.5 pixel, which for Chandra PSF  includes $\sim$ 98\% of counts from an on-axis point source (Figure 4.6 in the \textit{Chandra} manual at http://cxc.harvard.edu/proposer/POG/html/chap4.html). For the background region we used an annulus with inner and outer radii of 4.5 and 13 pixels, respectively, and excluding the 3.5 pixel circle centered at the position of the point source found just South of the supernova position by {\tt wavdetect}. In the source region we detected 131 counts, while the expected number of background counts (rescaled to the source region area) is $150.7$, giving net source counts of $-19.7\pm11.5$ counts. The negative signal in the source region is obviously caused by the complex non-uniform diffuse emission around the supernova position. Finally, we recomputed the average background level in the source aperture using the entire 13 pixel circle at the supernova position, while still excluding the point source to the South of the supernova position (see Figure \ref{Fig:erange_300:1000}), and obtained $148.2$ counts per 4.5 pixel aperture. From this value we compute the $3\sigma$ upper limit on the number of source counts as $\mu = 3\cdot \sqrt{148.2}=36.5$.

The upper limit of the luminosity of the hypothetical SSS in the 0.3-1 keV energy band was found using the formula
\begin{eqnarray}
 L_{X,UL} = 4 \pi \frac{ \mu \langle E_{\gamma} \rangle d^2}{\zeta} \label{Eq:L_X}
 \label{eq:ul}
\end{eqnarray}
where $\mu$ is the $3\sigma$ upper limit on the source count obtained above, $\langle E_{\gamma} \rangle$ is the average energy of the photons found from the absorbed \textsc{xspec} model for the assumed spectrum, $d$ is the distance to the SN and $\zeta$ is the value of the exposure map for the given spectrum at the position of the SN on the detector. Spectral weights needed to compute the exposure map, were calculated assuming an absorbed black body spectrum ({\tt xsphabs} and {\tt xsbbody} from \textsc{ciao}'s spectral fitting tool \textsc{sherpa}) for four values of effective temperature: 30, 50, 100 and 150 eV and with  $N_H=6.9\cdot10^{21}$ cm$^{-2}$ (see below).

The upper limit on the luminosity thus obtained needs to be corrected for the interstellar absorption. To this end we used  \textit{Chandra}'s \textsc{pimms} tool\footnote{http://cxc.harvard.edu/toolkit/pimms.jsp}. This yielded the unabsorbed luminosity of the source in the supersoft X-ray band (i.e. between 300 eV and 1 keV). Bolometric corrections were then applied to the unabsorbed luminosities according to each of the assumed effective temperatures to obtain the upper limits on the bolometric luminosity of the source. Our results are summarized in Table \ref{Table:Results}.

The absorbing neutral hydrogen column can be found from the reddening via the relation $N_{\mathrm{H}} = 2.21\cdot10^{21} A_V = 6.9 \cdot 10^{21} \cdot E(B-V)$, where $N_{\mathrm{H}}$ is the neutral hydrogen column in $\mathrm{cm}^{-2}$ \citep{Guver.Ozel.2009}, and $A_V$ is the visual extinction ($A_V=3.1\cdot E(B-V)$). At the time of writing, no clear consensus appears to exist as to the value of the reddening for SN2014J. Based on initial spectroscopy \citet{Cox.et.al.2014} gave a reddening value of $E(B-V) > 1$, whereas \citet{Kotak.2014} listed $E(B-V) \lesssim 0.8$. \citet{Altavilla.et.al.2014} give $B-V = 1.2$, which corresponds to $E(B-V) = 1.2$ if we assume an intrinsic $B-V\sim0$, as is customarily.  In the following, we conservatively  assumed a value of $N_{\mathrm{H}}=6.9\cdot10^{21}$ cm$^{-2}$ corresponding to the $E(B-V) = 1$. Note that this value is also consistent with the hydrogen column density determined from the X-ray spectral fitting of the spectrum of the diffuse emission around the supernova.

\begin{table*}
 \begin{minipage}{140mm}
 \caption{Upper limits of the supersoft X-ray and bolometric luminosities of the progenitor of SN2014J, assuming a blackbody SSS.}
 \centering
  \begin{tabular}{@{} c c c c c c c @{}}
  \hline
   $T_{\mathrm{eff}}$	& 3$\sigma$ UL source		& $\langle E_{\gamma} \rangle$	& $\zeta$		& 3$\sigma L_{X,\mathrm{UL}}$	& 3$\sigma L_{X,\mathrm{UL}}$	& 3$\sigma L_{\mathrm{bol},\mathrm{UL}}$ \\
			& counts (4.5 pixel		& 				&			& absorbed			& unabsorbed			& unabsorbed \\
   $[$eV] 		& radius region)		& [erg/count]			& [s$\cdot$cm$^{-2}$]	& [erg/s]			& [erg/s]			& [erg/s] \\
  \hline
    30			& 36.5				& 7.6$\cdot10^{-10}$		& 7.8$\cdot10^{7}$	& 5.2$\cdot10^{35}$		& 5.3$\cdot10^{39}$		& 5.5$\cdot10^{41}$ \\
    50			& 36.5				& 8.9$\cdot10^{-10}$		& 1.9$\cdot10^{8}$	& 4.0$\cdot10^{35}$		& 4.3$\cdot10^{38}$		& 3.1$\cdot10^{39}$ \\
    100			& 36.5				& 1.2$\cdot10^{-9}$		& 2.4$\cdot10^{8}$	& 2.7$\cdot10^{35}$		& 2.3$\cdot10^{37}$		& 3.8$\cdot10^{37}$ \\
    150			& 36.5				& 1.3$\cdot10^{-9}$		& 2.8$\cdot10^{8}$	& 2.5$\cdot10^{35}$		& 8.8$\cdot10^{36}$		& 1.2$\cdot10^{37}$ \\
    \hline
\end{tabular} \label{Table:Results}
\end{minipage}
\end{table*}

\begin{figure}
 \centering
  \includegraphics[width=1.0\linewidth]{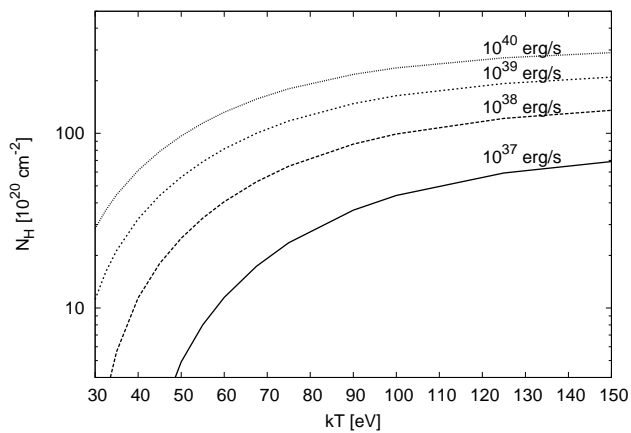} 
  \caption{3$\sigma$ upper limits on the bolometric luminosity (assuming a blackbody spectrum) of any putative supersoft X-ray source progenitor of SN2014J, for varying effective temperature and column density. The lines correspond to constant value of the upper limit on the $N_H-T_{\rm eff}$ plane. The curves are marked by the corresponding upper limit values.}\label{Fig:Teff_NH}
\end{figure}

To generalize our results to other values of $N_{\mathrm{H}}$, we show in  Figure \ref{Fig:Teff_NH} the behaviour of the upper limit values as a function of $T_{\mathrm{eff}}$ and $N_{\mathrm{H}}$.
In order to survey the large parameter space, we  made use of the \textsc{pimms} tool, instead of accurately computing the value of the exposure map at the supernova position.  As almost all of our pre-explosion observations were made during (501.33/828.11 ks) or before (304.93/828.11 ks) Chandra Cycle 10, and all but four (observations 378, 379, 380, and 1302, constituting only 33.6 out of the total 828 ksec) used the ACIS-S detector, we used the ACIS-S Cycle 10 instrument response in converting the count rate to energy flux. Comparison with our more precise calculations above shows that the accuracy of the \textsc{pimms}-based  calculation is better than $\sim$10\%. Note that we did not consider $N_{\rm{H}}$ $<$ $4 \cdot 10^{20}\rm{cm}^{-2}$, this being the Galactic foreground H I column density as given by \citet{Dickey.Lockman.1990}\footnote{http://heasarc.gsfc.nasa.gov/cgi-bin/Tools/w3nh/w3nh.pl}. 

For completeness, we also calculated upper limits on the X-ray luminosity in the 0.5-2 keV and 2-10 keV bands assuming a generic  $\Gamma$=2 power law spectrum. These upper limits are summarized in Table \ref{Table:Results_Xray_ULs}.

\begin{table*}
 \begin{minipage}{140mm}
 \caption{Upper limits of the soft and hard X-ray luminosities of the progenitor of SN2014J, assuming a $\Gamma=2$ power law.}
 \centering
  \begin{tabular}{@{} c c c c c c @{}}
  \hline
   photon		& 3$\sigma$ UL source		& $\langle E_{\gamma} \rangle$	& $\zeta$		& 3$\sigma L_{X,\mathrm{UL}}$	& 3$\sigma L_{X,\mathrm{UL}}$ \\
   energies		& counts (4.5 pixel 		& 				&			& absorbed			& unabsorbed \\
   $[$keV] 		& radius region)		& [erg/count]			& [s$\cdot$cm$^{-2}$]	& [erg/s]			& [erg/s] \\
  \hline
    0.5-2		& 57.1				& 2.2$\cdot10^{-9}$		& 4.5$\cdot10^{8}$	& 4.1$\cdot10^{35}$		& 1.5$\cdot10^{36}$ \\
    2-10		& 28.4				& 6.5$\cdot10^{-9}$		& 2.5$\cdot10^{8}$	& 1.1$\cdot10^{36}$		& 1.1$\cdot10^{36}$ \\
    \hline
\end{tabular} \label{Table:Results_Xray_ULs}
\end{minipage}
\end{table*}

\subsection{Post-supernova observations}
\begin{figure}
 \centering
  \includegraphics[width=1.0\linewidth]{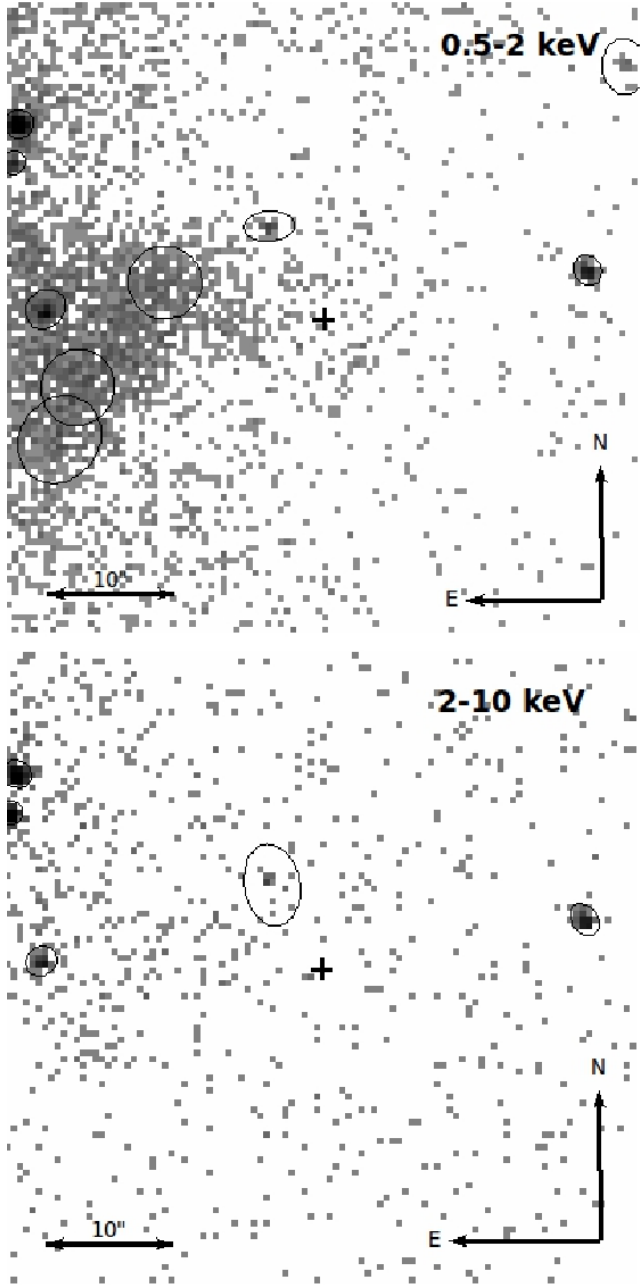}
  \caption{Part of \textit{Chandra} observation 16580, for soft and hard photon energies, showing the sources found by {\tt wavdetect} (thin ellipses). The crosses show the position of the supernova.}\label{Fig:soft_and_hard_energies_postSN}
\end{figure}

Using the same method as for the stacked pre-explosion observations, we calculated upper limits to the X-ray luminosity of the supernova after the explosion. As with the soft and hard bands mentioned above, we assume a generic powerlaw spectrum with the photon index of  $\Gamma=2$. Our results are summarised in Table \ref{Table:Results_Xray_ULs_post_SN}. Figure \ref{Fig:soft_and_hard_energies_postSN} shows the post-supenova observation images in the soft and hard bands.

\begin{table*}
 \begin{minipage}{140mm}
 \caption{Upper limits of the soft and hard X-ray luminosities of SN2014J, assuming a $\Gamma=2$ power law.}
 \centering
  \begin{tabular}{@{} c c c c c c @{}}
  \hline
   photon		& 3$\sigma$ UL source		& $\langle E_{\gamma} \rangle$	& $\zeta$		& 3$\sigma L_{X,\mathrm{UL}}$	& 3$\sigma L_{X,\mathrm{UL}}$ \\
   energies		& counts (4.5 pixel		& 				&			& absorbed			& unabsorbed \\
   $[$keV] 		& radius region)		& [erg/count]			& [s$\cdot$cm$^{-2}$]	& [erg/s]			& [erg/s] \\
  \hline
    0.5-2		& 12.4				& 2.2$\cdot10^{-9}$		& 2.4$\cdot10^{7}$	& 1.7$\cdot10^{36}$		& 6.1$\cdot10^{36}$ \\
    2-10		& 5.9				& 6.5$\cdot10^{-9}$		& 1.5$\cdot10^{7}$	& 3.8$\cdot10^{36}$		& 4.0$\cdot10^{36}$ \\
    \hline
\end{tabular} \label{Table:Results_Xray_ULs_post_SN}
\end{minipage}
\end{table*}

\section{Discussion} \label{Sect:Discussion}
With the upper limits presented in \citet{Nielsen.et.al.2012,Nielsen.et.al.2013b} the total number of nearby ($<$25 Mpc) type Ia supernovae for which pre-explosion \textit{Chandra} observations are available is now 15: SN2002cv, SN2003cg, SN2004W, SN2006X, SN2006dd, SN2006mr, SN2007gi, SN2007on, SN2007sr, SN2008fp, SN2011fe, SN2011iv, SN2012cu, SN2012fr and SN2014J. None of these show strong evidence of supersoft X-ray emissions from progenitors \citep{Voss.Nelemans.2008,Nielsen.et.al.2012,Nielsen.et.al.2013b}, although the case of SN2007on is ambiguous \citep{Roelofs.et.al.2008}.

For SN2014J, our results rule out a naked (i.e. one that is not obscured by local material) SSS progenitor of $L_{\mathrm{bol}} \gtrsim 3.8 \cdot 10^{37}$ erg/s if its effective temperature exceeds 100 eV. Therefore, if the accreting white dwarf in a single-degenerate type Ia supernova progenitor system has a photospheric radius comparable to the near-Chandrasekhar mass white dwarf radius, and is free of significant amounts of local obscuring material, then our results firmly rule out such a progenitor for SN2014J.

A similar conclusion has been reached for SN2007sr, SN2011fe and SN2012fr, and tentatively also for SN2006dd and SN2006mr (see Figure 2 in \citealt{Nielsen.et.al.2013b}). These results suggest that in these six objects either the progenitors are not single-degenerate systems, or their effective temperatures immediately before the explosion are lower than expected in the standard single-degenerate scenario.

Because of the considerable obscuring neutral hydrogen column, the Chandra observations do not provide significant constrains in the lower range of SSSs effective temperatures. The upper limit exceeds $\sim 10^{38}$ erg/s for temperatures below $\la 80$ eV.

\section{Conclusions} \label{Sect:Conclusion}
SN2014J in M82 is the nearest type Ia supernova in four decades, and the position of the progenitor system has been observed extensively by the \textit{Chandra} X-ray Telescope prior to the supernova explosion.

We examined the totality of pre-explosion observations from the \textit{Chandra} data archive and found no evidence of an X-ray progenitor for SN2014J. We calculated upper limits on the X-ray and bolometric luminosities of the progenitor (Tables \ref{Table:Results}, \ref{Table:Results_Xray_ULs}, Fig.\ref{Fig:Teff_NH}). Our results firmly rule out the progenitor being a single-degenerate system with a photospheric radius comparable to the near-Chandrasekhar mass white dwarf radius and mass accretion rate in the range where stable nuclear burning can occur. This is in agreement with results obtained for several other nearby type Ia supernovae with pre-explosion images in the \textit{Chandra} archive. Due to the presence of a large absorbing column we cannot rule out a low temperature supersoft X-ray progenitor with $kT_{\mathrm{eff}} \lesssim$ 80 eV.

We also examined the one existing \textit{Chandra} observation made after the supernova explosion, and found no evidence of X-ray emission on the position of the supernova. We calculated upper limits on the X-ray luminosity of SN2014J (Table \ref{Table:Results_Xray_ULs_post_SN}).

SN2014J is located in the center of a region of enhanced soft diffuse emission, roughly 200 pc across, which appears to be a supernova-inflated bubble around a region of recent star formation. The short cooling time of the gas, estimated from the X-ray spectral analysis, suggests a relatively young age of the bubble. If SN2014J is indeed located inside the bubble, it belongs to the prompt population of type Ia supernovae, with the delay time of the order of $\sim 50$ Myrs.

\label{lastpage}


\begin{thebibliography}{99}
\bibitem[\protect\citeauthoryear{Altavilla et al.}{2014}]{Altavilla.et.al.2014} Altavilla G. on behalf of the INAF-Astronomical Observatory of Bologna,ATEL\#5818
\bibitem[\protect\citeauthoryear{Cao et al.}{2014}]{Cao.et.al.2014} Cao, Y., Kasliwal, M.~M., McKay, A. \& Bradley, A., 2014, ATEL\#5786
\bibitem[\protect\citeauthoryear{Cox et al.}{2014}]{Cox.et.al.2014} Cox, N.~L.~J., Davis, P., Patat, F. \& van Winckel, H., 2014, ATEL\#5797
\bibitem[\protect\citeauthoryear{Dalcanton et al.}{2009}]{Dalcanton.et.al.2009} Dalcanton, J.~J., Williams, B.~F., Seth, A.~C.,	Dolphin, A., Holtzman, J., Rosema, K., Skillman, E.~D., Cole, A., Girardi, L., Gogarten, S.~M., Karachentsev, I.~D., Olsen, K., Weisz, D., Christensen, C., Freeman, K., Gilbert, K., Gallart, C., Harris, J., Hodge, P., de Jong, R.~S., Karachentseva, V., Mateo, M., Stetson, P.~B., Tavarez, M., Zaritsky, D., Governato, F. \& Quinn, T., 2009, ApJS, 183, 67
\bibitem[\protect\citeauthoryear{Dickey \& Lockman}{1990}]{Dickey.Lockman.1990} Dickey, J.~M., \& Lockman, F.~J.\ 1990, ARAA, 28, 215
\bibitem[\protect\citeauthoryear{Di Stefano}{2010}]{DiStefano.2010} Di Stefano, R., 2010, ApJ, 712, 728
\bibitem[\protect\citeauthoryear{Gilfanov \& Bogd{\'a}n}{2010}]{Gilfanov.Bogdan.2010} Gilfanov, M. \& Bogd{\'a}n, {\'A}., 2010, Nature, 463, 924
\bibitem[\protect\citeauthoryear{Greiner et al.}{1991}]{Greiner.et.al.1991} Greiner, J., Hasinger, G. \& Kahabka, P., 1991, A\&A, 246, L17
\bibitem[\protect\citeauthoryear{Greiner}{2000}]{Greiner.2000} Greiner, J., 2000, New Astron. 5, 137
\bibitem[\protect\citeauthoryear{G\"{u}ver \& \"{O}zel}{2009}]{Guver.Ozel.2009} G\"{u}ver, T. \& \"{O}zel, F., 2009, MNRAS, 400, 2050
\bibitem[\protect\citeauthoryear{van den Heuvel et al.}{1992}]{van.den.Heuvel.et.al.1992} van den Heuvel, E.~P.~J., Bhattacharya, D., Nomoto, K. and Rappaport, S.~A., 1992, 262, 97
\bibitem[\protect\citeauthoryear{Hillebrandt et al.}{2013}]{Hillebrandt.et.al.2013} Hillebrandt, W. et al., Kromer, M., R{\"o}pke, F.~K. \& Ruiter, A.~J., 2013, FrPhy, 8, 116
\bibitem[\protect\citeauthoryear{Iben \& Tutukov}{1984}]{Iben.Tutukov.1984} Iben, I. and Tutukov, A.~V., 1984, ApJS, 54, 335
\bibitem[\protect\citeauthoryear{Johansson et al.}{2014}]{Johansson.et.al.2014} Johansson, J., Woods, T.~E., Gilfanov, M., Sarzi, M., Chen, Y.-M. \& Oh, K., 2014, arXiv e-prints \#1401.1344
\bibitem[\protect\citeauthoryear{Kaastra \& Mewe}{1993}]{Kaastra.Mewe.1993} Kaastra, J.~S. \& Mewe, R., 1993, AAPS, 97, 443
\bibitem[\protect\citeauthoryear{Kahabka \& van den Heuvel}{1997}]{Kahabka.van.den.Heuvel.1997} Kahabka, P. and van den Heuvel, E.~P.~J., 1997, ARA\&A, 35, 69
\bibitem[\protect\citeauthoryear{Kotak}{2014}]{Kotak.2014} Kotak, R., 2014, ATEL\#5816
\bibitem[\protect\citeauthoryear{Liedahl et al.}{1995}]{Liedahl.et.al.1995} Liedahl, D.~A., Osterheld, A.~L. \& Goldstein, W.~H., 1995, ApJL, 438, L115
\bibitem[\protect\citeauthoryear{Mac Low \& McCray}{1988}]{Mac.Low.McCray.1988} Mac Low, M.-M. \& McCray, R., 1988, ApJ, 324, 776
\bibitem[\protect\citeauthoryear{Maksym et al.}{2014}]{Maksym.et.al.2014} Maksym, W.~P., Irwin, J.~A., Keel, W.~C., Burke, D \& K. Schawinski, K. ATEL\#5798
\bibitem[\protect\citeauthoryear{Mannucci et al.}{2005}]{Mannucci.et.al.2005} Mannucci, F., Della Valle, M., Panagia, N., Cappellaro, E., Cresci, G. Maiolino, R, Petrosian, A. \& Turatto, M., 2005, A\&A 433, 807
\bibitem[\protect\citeauthoryear{Maoz et al.}{2013}]{Maoz.et.al.2013} Maoz, D., Mannucci, F. \& Nelemans, G., 2013, arXiv e-prints \#1312.0628
\bibitem[\protect\citeauthoryear{Margutti et al.}{2014}]{Margutti.et.al.2014} Margutti, R., Soderberg, A., Kamble, A., Zauderer, A., Milisavljevic, D., Parrent, J. \& Chomiuk, L., 2014, ATEL\#5851
\bibitem[\protect\citeauthoryear{McCray \& Kafatos}{1987}]{McCray.Kafatos.1987} McCray, R. \& Kafatos, M., 1987, ApJ, 317, 190
\bibitem[\protect\citeauthoryear{Nielsen et al.}{2012}]{Nielsen.et.al.2012} Nielsen, M.~T.~B., Voss, R., \& Nelemans, G., 2012, MNRAS, 426, 2668
\bibitem[\protect\citeauthoryear{Nielsen et al.}{2013a}]{Nielsen.et.al.2013a} Nielsen, M.~T.~B., Dominik, C., Nelemans, G. \& Voss, R., 2013a, A\&A, 549, A32
\bibitem[\protect\citeauthoryear{Nielsen et al.}{2013b}]{Nielsen.et.al.2013b} Nielsen, M.~T.~B., Voss, R. \& Nelemans, G., 2013b, MNRAS, 435, 187
\bibitem[\protect\citeauthoryear{Nielsen et al.}{2014a}]{Nielsen.et.al.2014a} Nielsen, M.~T.~B., Nelemans, G., Voss, R. \& Toonen, S., A\&A, 2014 (accepted)
\bibitem[\protect\citeauthoryear{Nielsen et al.}{2014b}]{Nielsen.et.al.2014b} Nielsen, M.~T.~B., Gilfanov, M., Woods, T.~E. \& Nelemans, G., 2014b, ATEL\#5799
\bibitem[\protect\citeauthoryear{Perlmutter et al.}{1999}]{Perlmutter.et.al.1999} Perlmutter, S. et al., 1999, ApJ, 517, 565
\bibitem[\protect\citeauthoryear{Philips}{1993}]{Philips.1993} Philips, M.~M., 1993, ApJ, 413, L105 
\bibitem[\protect\citeauthoryear{Riess et al.}{1998}]{Riess.et.al.1998} Riess, A.~G. et al., 1998, AJ, 116, 1009
\bibitem[\protect\citeauthoryear{Roelofs et al.}{2008}]{Roelofs.et.al.2008} Roelofs, G., Bassa, C., Voss, R. and Nelemans, G., 2008, 391, 290
\bibitem[\protect\citeauthoryear{Scannapieco \& Bildsten}{2005}]{Scannapieco.Bildsten.2005} Scannapieco, E. \& Bildsten, L., 2005, ApJ, 629, L85
\bibitem[\protect\citeauthoryear{STSI}{2014}]{STSI.2014} STSI - Space Telescope Science Institute, 2014, ATEL\#5821
\bibitem[\protect\citeauthoryear{Toonen et al.}{2012}]{Toonen.et.al.2012} Toonen, S, Nelemans, G. \& Portegies-Zwart, S., 2012, A\&A, 546, A70
\bibitem[\protect\citeauthoryear{Voss \& Nelemans}{2008}]{Voss.Nelemans.2008} Voss, R. and Nelemans, G., 2008, Nature, 451, 802
\bibitem[\protect\citeauthoryear{Webbink}{1984}]{Webbink.1984} Webbink, R.~F., 1984, ApJ, 277, 355
\bibitem[\protect\citeauthoryear{Whelan \& Iben}{1973}]{Whelan.Iben.1973} Whelan, J. and Iben, I.~J., 1973, ApJ, 186, 1007
\bibitem[\protect\citeauthoryear{Woods \& Gilfanov}{2013}]{Woods.Gilfanov.2013} Woods, T.~E., \& Gilfanov, M., 2013, MNRAS, 432, 1640
\end{thebibliography}
\end{document}